\documentclass[aps,prl,twocolumn,showpacs,amsmath,amssymb,superscriptaddress]{revtex4}
\usepackage{graphicx}
\usepackage{amssymb}
\usepackage{natbib}
\usepackage{color}

\newcommand{\beq}{\begin{eqnarray}}
\newcommand{\eeq}{\end{eqnarray}}

\begin{document}

\title{Structural and magnetic phase transitions near optimal superconductivity in
 BaFe$_2${(As$_{1-x}$P$_x$)}$_2$ }
\author{Ding Hu}
\affiliation{Institute of Physics, Chinese Academy of Sciences, Beijing
100190, China}
\author{Xingye Lu}
\affiliation{Institute of Physics, Chinese Academy of Sciences, Beijing
100190, China}
\author{Wenliang Zhang}
\affiliation{Institute of Physics, Chinese Academy of Sciences, Beijing
100190, China}
\author{Huiqian Luo}
\affiliation{Institute of Physics, Chinese Academy of Sciences, Beijing
100190, China}
\author{Shiliang Li}
\affiliation{Institute of Physics, Chinese Academy of Sciences, Beijing
100190, China}
\affiliation{Collaborative Innovation Center of Quantum Matter, Beijing, China}
\author{Peipei Wang}
\affiliation{Institute of Physics, Chinese Academy of Sciences, Beijing
100190, China}
\author{Genfu Chen}
\email{gfchen@iphy.ac.cn}
\affiliation{Institute of Physics, Chinese Academy of Sciences, Beijing
100190, China}
\author{Fei Han}
\affiliation{Materials Science Division,
Argonne National Laboratory, Argonne, Illinois 60439, USA}
\author{Shree R. Banjara}
\affiliation{Ames Laboratory, US DOE, Ames, IA, 50011, USA}
\affiliation{Department of Physics and Astronomy, Iowa State University, Ames, IA, 50011, USA}
\author{A. Sapkota}
\affiliation{Ames Laboratory, US DOE, Ames, IA, 50011, USA}
\affiliation{Department of Physics and Astronomy, Iowa State University, Ames, IA, 50011, USA}
\author{A. Kreyssig}
\affiliation{Ames Laboratory, US DOE, Ames, IA, 50011, USA}
\affiliation{Department of Physics and Astronomy, Iowa State University, Ames, IA, 50011, USA}
\author{A. I. Goldman}
\affiliation{Ames Laboratory, US DOE, Ames, IA, 50011, USA}
\affiliation{Department of Physics and Astronomy, Iowa State University, Ames, IA, 50011, USA}
\author{Z. Yamani}
\affiliation{Canadian Neutron Beam Centre, National Research Council, Chalk River, Ontario, K0J 1P0 Canada
}
\author{Christof Niedermayer}
\affiliation{Laboratory for Neutron Scattering, Paul Scherrer Institut, CH-5232 Villigen, Switzerland
}
\author{Markos Skoulatos}
\affiliation{Laboratory for Neutron Scattering, Paul Scherrer Institut, CH-5232 Villigen, Switzerland
}
\author{Robert Georgii}
\affiliation{Heinz Maier-Leibnitz Zentrum, Technische Universit$\ddot{a}$t M$\ddot{u}$nchen, D-85748 Garching, Germany
}
\author{T. Keller}
\affiliation{Max-Planck-Institut f$\ddot{u}$r Festk$\ddot{o}$rperforschung, Heisenbergstrasse 1, D-70569 Stuttgart, Germany}
\affiliation{Max Planck Society Outstation at the Forschungsneutronenquelle Heinz Maier-Leibnitz (MLZ), D-85747 Garching, Germany}
\author{Pengshuai Wang}
\affiliation{Department of Physics, Renmin University of China, Beijing 100872, China
}
\author{Weiqiang Yu}
\affiliation{Department of Physics, Renmin University of China, Beijing 100872, China
}

\author{Pengcheng Dai}
\email{pdai@rice.edu}
\affiliation{Department of Physics and Astronomy, Rice University, Houston, Texas 77005, USA
}
\affiliation{Institute of Physics, Chinese Academy of Sciences, Beijing
100190, China}
\date{\today}

\begin{abstract}
We use nuclear magnetic resonance (NMR), high-resolution x-ray and neutron scattering to study
structural and magnetic phase transitions in phosphorus-doped
BaFe$_2${(As$_{1-x}$P$_x$)}$_2$.  Previous transport, NMR, specific heat, and magnetic penetration depth
measurements have provided compelling evidence for the presence
of a quantum critical point (QCP) near optimal superconductivity at $x=0.3$.
However, we show that the  tetragonal-to-orthorhombic structural ($T_s$) and paramagnetic to
antiferromagnetic (AF, $T_N$) transitions in BaFe$_2${(As$_{1-x}$P$_x$)}$_2$ are always coupled and approach to
$T_N\approx T_s \ge T_c$ ($\approx 29$ K) for $x=0.29$ before vanishing abruptly for $x\ge 0.3$.
These results suggest that AF order in BaFe$_2${(As$_{1-x}$P$_x$)}$_2$ disappears in a weakly first order fashion near optimal superconductivity,
 much like the electron-doped iron pnictides with an avoided QCP.
\end{abstract}

\maketitle

A determination of the structural and magnetic phase diagrams
in different classes of iron pnictide superconductors will form the basis from which a microscopic theory of superconductivity
can be established \cite{kamihara,cruz,qhuang,mgkim11,dai}.  The parent compound of iron pnictide superconductors
such as
BaFe$_2$As$_2$ exhibits a tetragonal-to-orthorhombic structural transition at temperature $T_s$ and
then orders antiferromagnetically below  $T_N$ with a collinear antiferromagnetic (AF)  structure
[Fig. 1(a)] \cite{qhuang,mgkim11}. Upon hole-doping via partially replacing Ba by K or Na \cite{Rotter,Gil},
the structural and magnetic phase transition temperatures in
Ba$_{1-x}A_x$Fe$_2$As$_2$ ($A=$ K, Na) decreases simultaneously with increasing $x$ and form a small pocket of a magnetic tetragonal phase with the $c$-axis aligned moment before disappearing abruptly near optimal superconductivity \cite{Avci2012,Avci2014,Waber,bohmer}.
For electron-doped Ba(Fe$_{1-x}T_x$)$_2$As$_2$ ($T=$Co,Ni), transport \cite{canfield,fisher}, muon
spin relaxation ($\mu$SR) \cite{bernhard},
nuclear magnetic resonance (NMR) \cite{ning,zhou2013,dioguardi}, x-ray and neutron
scattering experiments \cite{clester,nandi,dkpratt11,hqluo,xylu,xylu14} have revealed that
the structural and magnetic phase transition
temperatures decrease and separate
with increasing $x$ \cite{clester,nandi,dkpratt11,hqluo,xylu,xylu14}.  However, instead of
a gradual suppression to zero temperature
near optimal superconductivity as expected for a magnetic quantum critical point (QCP) \cite{ning,zhou2013},
the AF order for Ba(Fe$_{1-x}T_x$)$_2$As$_2$
 near optimal superconductivity actually occurs around 30 K ($>T_c$) and forms a short-range
incommensurate magnetic phase which
competes with superconductivity and disappears in the weakly first order fashion,
thus avoiding the expected magnetic QCP \cite{dkpratt11,hqluo,xylu,xylu14}.

Although a QCP may be avoided in electron-doped Ba(Fe$_{1-x}T_x$)$_2$As$_2$ due to disorder and
impurity scattering in the FeAs plane
induced by Co and Ni substitution,  phosphorus-doped
BaFe$_2${(As$_{1-x}$P$_x$)}$_2$ provides an alternative system to achieve
a QCP since substitution of As by the isovalent P suppresses the static AF order
and induces superconductivity without appreciable impurity scattering \cite{eabrahams11,SJiang,Shishido10,Beek10}.
Indeed, experimental evidence for the presence of a QCP at $x=0.3$ in BaFe$_2${(As$_{1-x}$P$_x$)}$_2$
has been mounting, including the linear temperature dependence of the resistivity \cite{Kasahara10},
an increase in the effective electron mass seen from the de Haas-van Alphen \cite{Shishido10},
magnetic penetration depth \cite{Hashimoto12,Shibauchi14}, heat capacity \cite{Walmsley}, and normal state
transport measurements in samples where superconductivity has been suppressed by a magnetic field \cite{Analytis14}.
Although these results, as well as
NMR measurements \cite{Nakai10}, indicate a QCP originating from the suppression of the
static AF order near $x=0.3$,
recent neutron powder diffraction experiments directly measuring $T_s$ and $T_N$ in BaFe$_2${(As$_{1-x}$P$_x$)}$_2$ as a function
of $x$ suggest that structural quantum criticality cannot exist at compositions higher
than $x = 0.28$ \cite{allred}.  Furthermore, the structural and magnetic phase transitions at all studied P-doping levels
are first order and occur
simultaneously within the sensitivity of the measurements ($\sim$0.5 K), thus casting doubt
on the presence of a QCP \cite{allred}.  While these results are interesting, they were
carried out on powder samples, and thus are not sensitive enough to the weak structural/magnetic order
to allow a conclusive determination on the nature of the
structural and AF phase transitions near optimal superconductivity.

\begin{figure}[t]
\includegraphics[scale=.65]{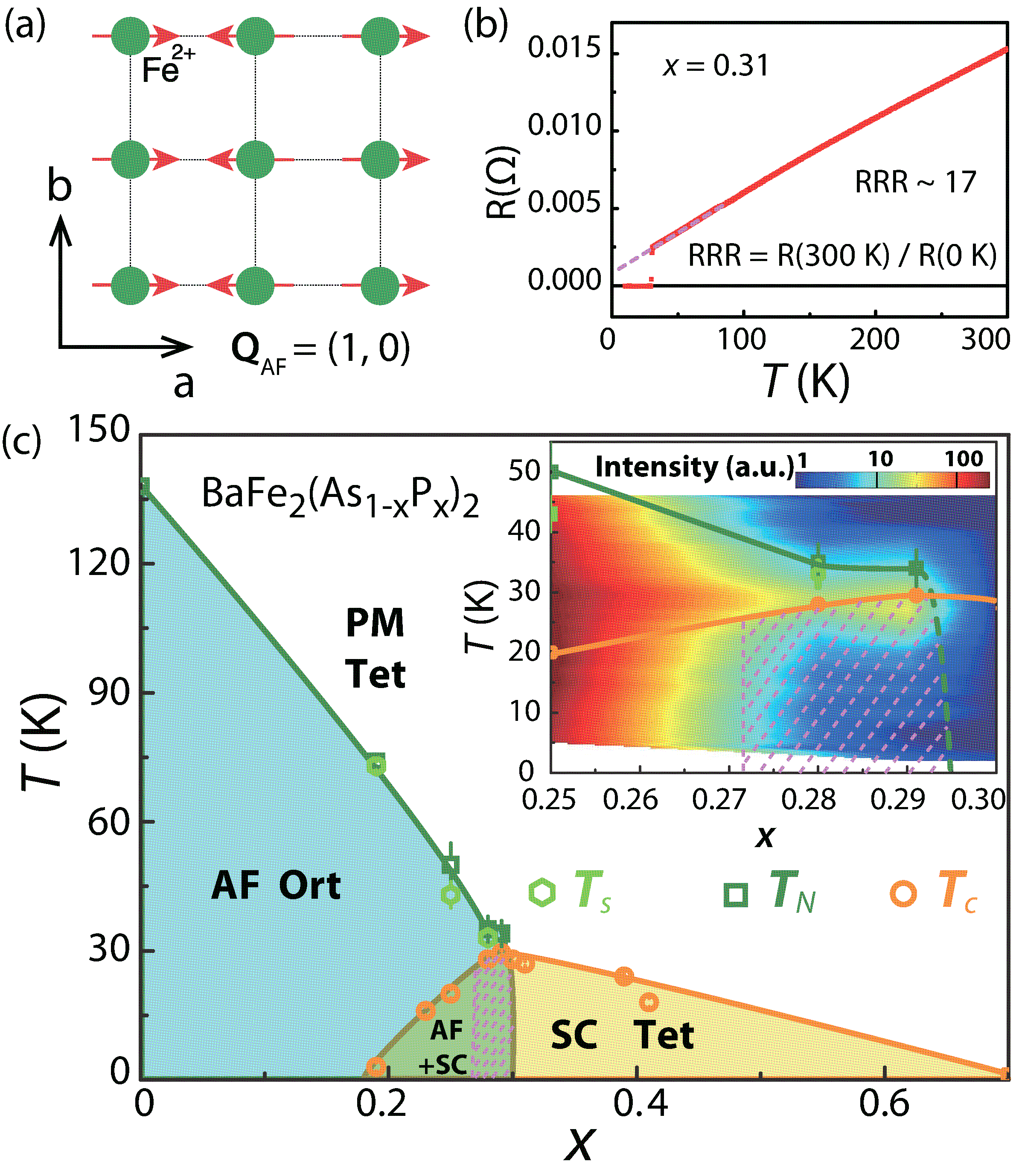}
\caption{
(a) The AF ordered phase of BaFe$_2$(As$_{1-x}$P$_x$)$_2$, where
the magnetic Bragg peaks occur at ${\bf Q}_{\rm AF}=(1,0,L)$ ($L=1,3,\cdots$) positions.
(b) Temperature dependence of the resistance for the $x=0.31$ sample,
where $RRR=R(300\ {\rm K})/R(0\ {\rm K})\sim 17$.  In previous work on similar P-doped samples,
$RRR\sim 13$ \cite{Kasahara10}.
(c) The phase diagram of BaFe$_2$(As$_{1-x}$P$_x$)$_2$, where the Ort, Tet, and SC are orthorhombic, tetragonal, and superconductivity phases, respectively.
The inset shows the expanded view of the
P-concentration dependence of $T_s$, $T_N$ and, $T_c$ near optimal superconductivity.
The color bar represents the
 temperature and doping dependence of the normalized magnetic Bragg peak intensity.
The dashed region indicates the mesoscopic coexisting AF and SC phases.
}
\end{figure}

In this Letter, we report systematic transport, NMR, x-ray and neutron scattering studies of
BaFe$_2$(As$_{1-x}$P$_x$)$_2$ single crystals focused on determining the
P-doping evolution of the structural and magnetic phase transitions near $x=0.3$.  While our data for
$x\le 0.25$ are consistent with the earlier results obtained from powder samples \cite{allred},
we find that nearly simultaneous structural and magnetic transitions in single crystals of BaFe$_2${(As$_{1-x}$P$_x$)}$_2$
occur at $T_s\approx T_N\ge T_c=29$ K for $x=0.28$ and 0.29 (near optimal doping) and disappear
suddenly at $x\ge 0.3$.  While superconductivity dramatically suppresses the static AF
order and lattice orthorhombicity below $T_c$ for $x=0.28$ and 0.29,
the collinear static AF order persists in the superconducting state.
Our neutron spin echo and NMR measurements on the $x=0.29$ sample
reveal that only part of the sample is magnetically ordered, suggesting its
mesoscopic coexistence with superconductivity. Therefore, in spite of reduced impurity scattering,
P-doped BaFe$_2$As$_2$ has remarkable similarities in the phase diagram to that of
electron-doped Ba(Fe$_{1-x}T_x$)$_2$As$_2$ iron pnictides with an avoided QCP.

\begin{figure}[t]
\includegraphics[scale=.7]{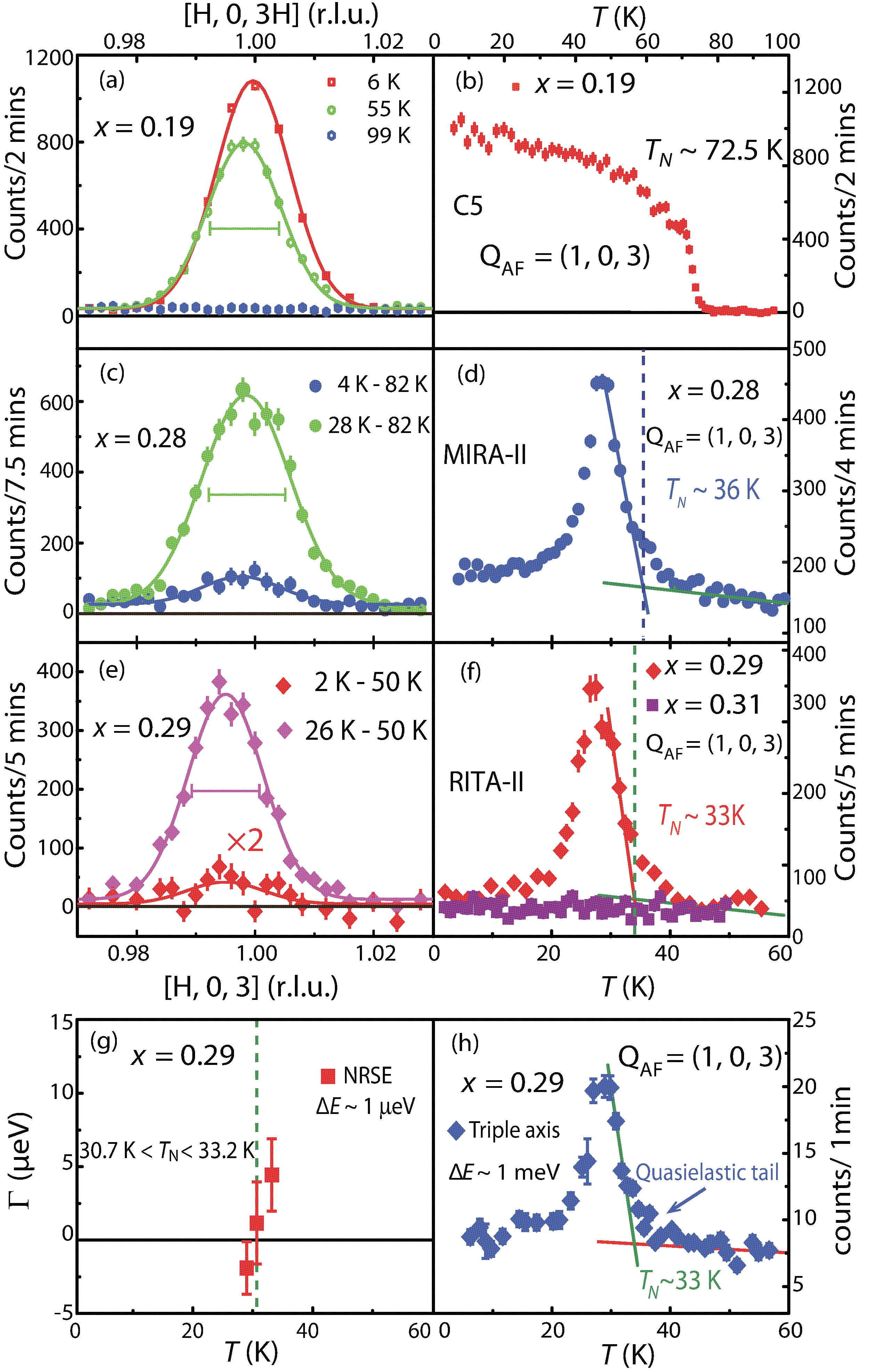}
\caption{(a,c,e)
Wave vector scans along the $[H,0,3]$ direction at different temperatures
for $x=0.19, 0.28$, 0.29, and 0.31, respectively.  Horizontal bars indicate instrumental resolution.
(b,d,f) Temperature dependence of the magnetic scattering at ${\bf Q}_{\rm AF}=(1,0,3)$
for $x=0.19, 0.28$, and 0.29, respectively.
(g) NRSE measurement of temperature dependence of the energy width ($\Gamma$ is the HWHM of scattering function and zero indicates
instrumental resolution limited) at ${\bf Q}_{\rm AF}=(1,0,3)$ for $x=0.29$. (h) The magnetic order parameters
from the normal triple-axis measurement on the same sample.
 }
 \end{figure}

We have carried out systematic neutron scattering experiments on BaFe$_2$(As$_{1-x}$P$_x$)$_2$ with
$x=0.19, 0.25, 0.28, 0.29, 0.30$, and 0.31 \cite{sample} using the C5, RITA-II, and MIRA
triple-axis spectrometers at the Canadian Neutron Beam center, Paul Scherrer Institute, and
Heinz Maier-Leibnitz Zentrum (MLZ), respectively.  We have also carried out neutron
resonance spin echo (NRSE) measurements on the $x=0.29$ sample using TRISP triple-axis
spectrometer at MLZ \cite{keller}. Finally, we have performed high-resolution
x-ray diffraction experiments on identical samples at Ames laboratory and Advanced Photon
Source, Argonne National Laboratory \cite{supplementary}.
Our single crystals were grown using Ba$_2$As$_2$/Ba$_2$P$_3$ self-flux method and
the chemical compositions are determined by inductively coupled plasma
analysis with 1\% accuracy \cite{sample}.
We define the wave vector
${\bf Q}$ at $(q_x, q_y, q_z)$ as $(H, K, L) = (q_xa/2\pi, q_yb/2\pi, q_zc/2\pi)$ reciprocal lattice units (rlu)
using the orthorhombic unit cell suitable for the AF
ordered phase of iron pnictides, where $a\approx b\approx 5.6$ \AA\ and $c=12.9$ \AA.
Figure 1(b) shows temperature dependence of the resistivity for $x=0.31$ sample,
confirming the high quality of our single crystals \cite{Kasahara10}.

\begin{figure}[t]
\includegraphics[scale=.75]{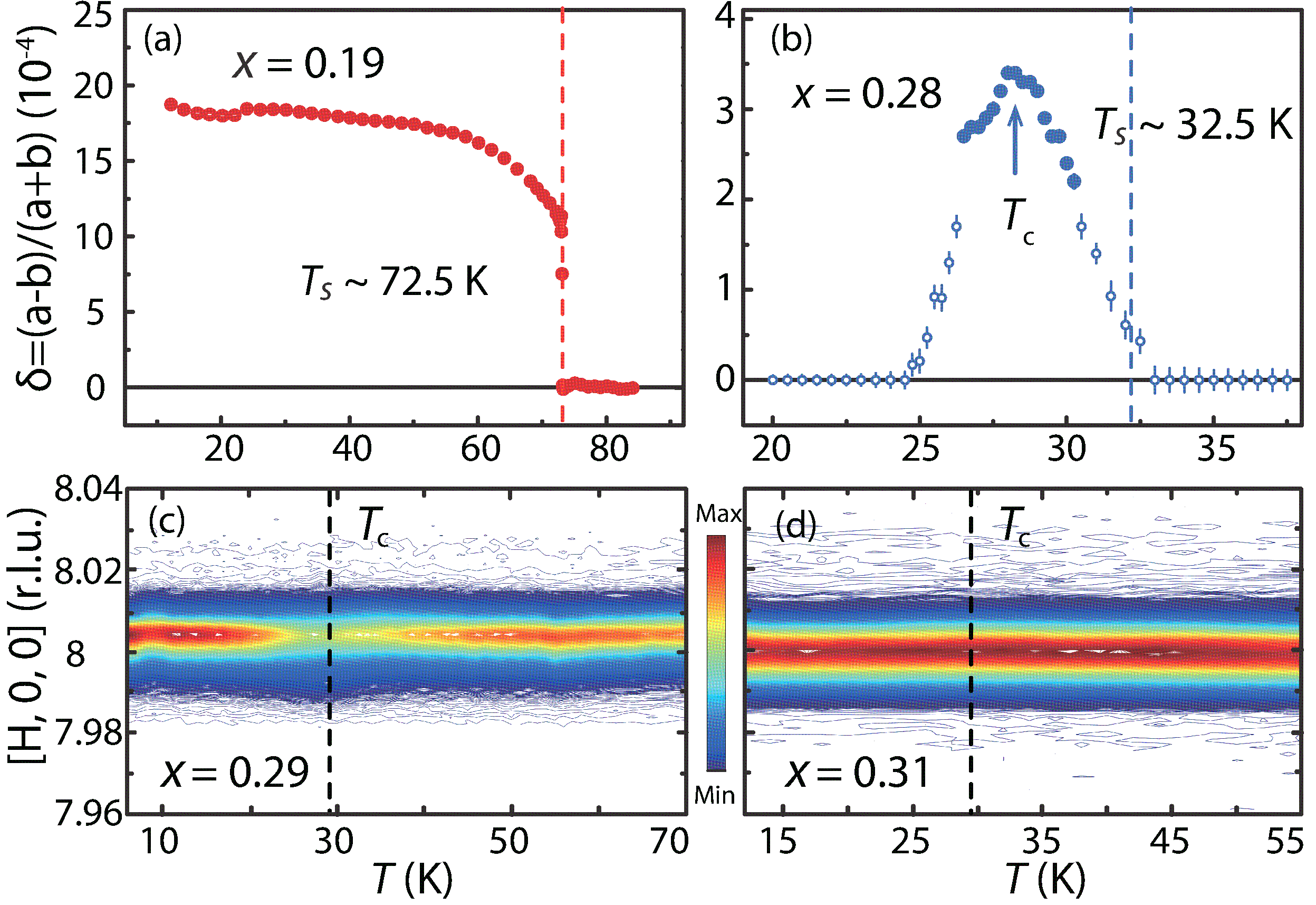}
\caption{Temperature evolution of $\delta$ for (a) $x=0.19$ and
(b) $x=0.28$ samples.  The solid circles indicate X-ray data where clear orthorhombic lattice
distortions are seen.  The open circles are data where one can only see peak broadening due to
orthorhombic lattice distortion. Temperature dependence of the $[H,0,0]$
scans for (c) $x=0.29$ and (d) $x=0.31$.  The vertical color
bar indicates X-ray scattering intensity.
The data were collected while warming system from base
temperature to a temperature well above $T_s$.
 }
\end{figure}

Figure 1(c) summarizes the phase diagram of BaFe$_2${(As$_{1-x}$P$_x$)}$_2$ as
determined from our experiments. Similar to previous
work on powder samples with $x\le 0.25$ \cite{allred}, we find that the structural and
AF phase transitions for single crystals of $x=0.19, 0.28$, and 0.29 occur simultaneously
within the sensitivity of our measurements ($\sim$1 K). On approaching optimal superconductivity as $x\rightarrow 0.3$,
the structural and magnetic phase transition temperatures are suppressed to $T_s\approx T_N\approx 30$ K for
$x=0.28, 0.29$ and then vanish suddenly for $x=0.3, 0.31$ as shown in the inset of Fig. 1(c).
Although superconductivity dramatically suppresses the lattice orthorhombicity and static AF order
in $x=0.28, 0.29$, there are still remnant
static AF order at temperatures well below $T_c$.
However, we find no evidence of static AF order and lattice orthorhombicity for
$x=0.3$ and 0.31 at all temperatures.
Since our NMR measurements on the $x=0.29$ sample suggest that the magnetic
order takes place in about $\sim$50\% of the volume fraction,
the coupled $T_s$ and $T_N$ AF phase in BaFe$_2${(As$_{1-x}$P$_x$)}$_2$ becomes a homogeneous superconducting phase in the weakly
first order fashion, separated by a phase with coexisting AF clusters and superconductivity [dashed region in Fig. 1(c)].

To establish the phase diagram in Fig. 1(c), we first present neutron scattering data aimed at
determining the N$\rm \acute{e}$el temperatures of BaFe$_2${(As$_{1-x}$P$_x$)}$_2$.
Figure 2(a) shows scans along the $[H,0,3H]$ direction at different temperatures for the $x=0.19$ sample.
The instrumental resolution limited peak centered at ${\bf Q}_{\rm AF}=(1,0,3)$
disappears at 99 K above $T_N$ [Fig. 2(a)]. Figure 2(b) shows the temperature dependence of the scattering
at ${\bf Q}_{\rm AF}=(1,0,3)$, which reveals a rather sudden change at $T_N=72.5\pm 1$ K consistent with the
first order nature of the magnetic transition \cite{allred}.
Figure 2(c) plots
$[H,0,0]$ scans through the $(1,0,3)$ Bragg peak
showing the temperature differences between 28 K (4 K) and 82 K for the $x=0.28$ sample.  There is a clear
resolution-limited peak centered
at $(1,0,3)$ at 28 K indicative of the static AF order, and the scattering is suppressed but not eliminated
at 4 K.  Figure 2(d) shows the temperature dependence of the scattering
at $(1,0,3)$, revealing a continuously increasing magnetic order parameter near $T_N$ and a dramatic
suppression of the magnetic intensity below $T_c$.  Figures 2(e) and 2(f) indicate that the magnetic order in
the $x=0.29$ sample behaves similar to that of the $x=0.28$ crystal without much reduction in $T_N$.
On increasing the doping levels to $x=0.3$ \cite{supplementary} and 0.31 [Fig. 2(f)], we find no evidence of magnetic order
above 2 K.  Given that the magnetic order parameters near $T_N$ for the $x=0.28, 0.29$ samples
look remarkably like those of the spin cluster phase in electron-doped
Ba(Fe$_{1-x}T_x$)$_2$As$_2$ near optimal superconductivity \cite{xylu,xylu14}, we have carried out
additional neutron scattering
measurements on the $x=0.29$ sample using TRISP, which can operate as a normal
thermal triple-axis spectrometer with instrumental energy resolution of $\Delta E \approx 1$ meV and a
NRSE triple-axis spectrometer with $\Delta E \approx 1\ \mu$eV \cite{keller}.
Fig. 2(h) shows the triple-axis mode data which reproduces the results in Fig. 2(f).  However, identical
measurements using NRSE mode reveals that the magnetic scattering above 30.7 K is quasielastic and
the spins of the system freeze below 30.7 K on a time scale of $\tau\sim \hbar/\Delta E \approx 6.6\times
10^{-10}$ s \cite{xylu14}.  This spin freezing temperature is almost identical to those of nearly optimally
electron-doped Ba(Fe$_{1-x}T_x$)$_2$As$_2$ \cite{hqluo,xylu,xylu14}.

 \begin{figure}[t]
\includegraphics[scale=.6]{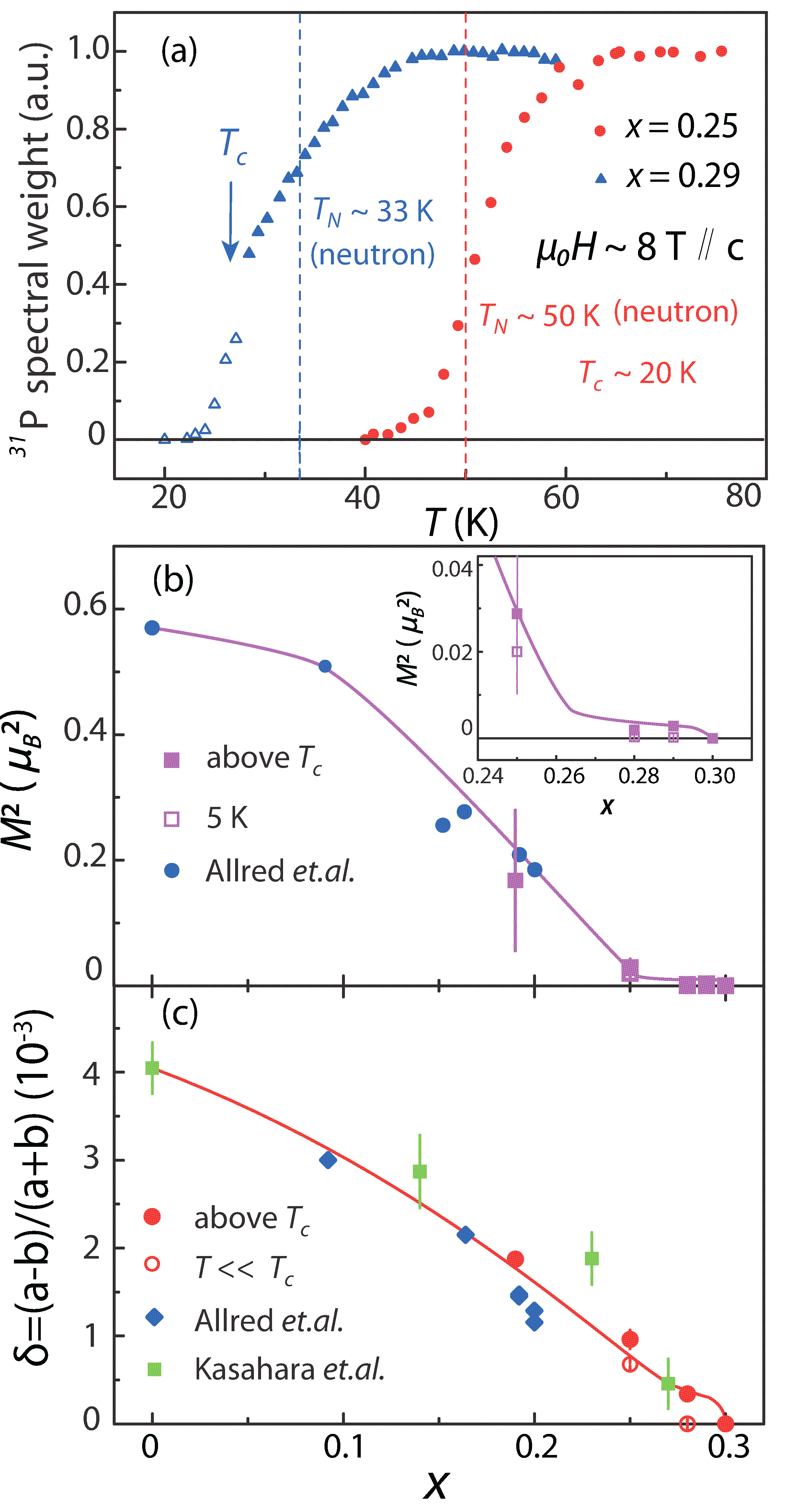}
\caption{(a) Temperature dependence of the paramagnetic spectral weight for $x=0.25$ and 0.29 samples from NMR measurements.
For $x=0.25$, there are no paramagnetic phase below 40 K, suggesting a fully magnetic ordered
phase.  At $T_c$ of the $x=0.29$ sample, there are still 50\% paramagnetic phase suggesting
the presence of magnetic signal
outside of the radio frequency window of the NMR measurement. The spectral weight loss below $T_c$ is due to superconductivity.
The vertical dashed lines mark $T_N$ determined from neutron scattering.
(b) The P-doping dependence of the $M^2$ estimated
from normalizing the magnetic Bragg intensity to weak nuclear peaks
assuming 100\% magnetically ordered phase.
The blue solid circles are from \cite{allred}.  The P-doping levels for different experiments are
normalized by their $T_N$'s.
 The inset shows the expanded view of $M^2$ around optimal doping above (solid squares) and
below (open squares) $T_c$. (c)
The P-doping dependence of $\delta$, where the
blue diamonds and green squares are from Refs. \cite{allred} and \cite{Kasahara10}, respectively.
For samples near optimal superconductivity, the fill and open red circles are $\delta$ above
and below $T_c$, respectively.
 }
\end{figure}

Figure 3 summarizes the key results of our x-ray
scattering measurements carried out on identical samples as those used
for neutron scattering experiments.
To facilitate quantitative comparison with the results on
Ba(Fe$_{1-x}T_x$)$_2$As$_2$, we define the
lattice orthorhombicity $\delta=(a-b)/(a+b)$ \cite{nandi,xylu}.  Figure 3(a) shows the temperature dependence of $\delta$ for
BaFe$_2${(As$_{1-x}$P$_x$)}$_2$ with $x=0.19$, obtained by fitting the
two Gaussian peaks in longitudinal scans along the $(8,0,0)$ nuclear Bragg peak \cite{supplementary}. We find that
the lattice orthorhombicity $\delta$ exhibits a first order like jump below $T_s=72.5$ K
consistent with previous neutron scattering results \cite{allred,supplementary}. We also note
that the lattice distortion value of $\delta \approx 17\times 10^{-4}$ is
similar to those of Ba(Fe$_{1-x}T_x$)$_2$As$_2$ with $T_s\approx 70$ K \cite{nandi,xylu}.

Figure 3(b) shows the temperature dependence of $\delta$ estimated for the $x=0.28$ sample.
In contrast to the $x=0.19$ sample, we only find clear evidence of lattice
orthorhombicity in the temperature region of $26\le T \le 32.5$ K [filled circles in
Fig. 3(b)] \cite{supplementary}.
The open symbols represent $\delta$ estimated from the enlarged half width of single peak fits
\cite{supplementary}.  Although the data suggests a re-entrant
tetragonal phase and vanishing lattice orthorhombicity at low temperature,
the presence of weak collinear AF order seen by neutron scattering
[Figs. 2(c) and 2(d)] indicates that the AF ordered parts of the sample
 should still have orthorhombic lattice distortion \cite{nandi,xylu}.
Figure 3(c) and 3(d) shows temperature dependence of the longitudinal scans along the
$[H,0,0]$ direction for the $x=0.29$ and 0.31 samples, respectively.   While the lattice distortion in
the $x=0.29$ sample behaves similarly as that of the $x=0.28$ crystal, there are no
observable lattice distortions in the probed temperature range for the $x=0.31$ sample.

To further test the nature of the magnetic ordered state in BaFe$_2${(As$_{1-x}$P$_x$)}$_2$,
we have carried out $^{31}$P NMR measurements under a 8-T $c$-axis aligned
magnetic field \cite{supplementary}.  Figure 4(a) shows the temperature dependence of the
integrated spectral weight of the paramagnetic signal, normalized by the Boltzmann factor, for single crystals
 with $x=0.25$ and 0.29.
For $x=0.25$, the paramagnetic spectral weight starts to drop below 60 K and reaches zero at 40 K,
suggesting a fully ordered magnetic state below 40 K. For $x=0.29$, the paramagnetic
to AF transition becomes much broader, and the magnetic ordered phase is estimated to be about 50\%
at $T_c=28.5$ K. Upon further cooling, the paramagnetic spectral weight drops dramatically below $T_c$ because of radio frequency screening.
We find that the lost NMR spectral weight above $T_c$  is not recovered at other frequencies,
suggesting that the magnetic ordered phase does not take full volume of the sample similar to the spin-glass
state of Ba(Fe$_{1-x}T_x$)$_2$As$_2$ \cite{hqluo,xylu,xylu14}.

Figure 4(b) shows the P-doping dependence of the ordered moment squared $M^2$ in BaFe$_2${(As$_{1-x}$P$_x$)}$_2$ including data
from Ref. \cite{allred}.  While $M^2$ gradually decreases
with increasing $x$ for $x\le 0.25$, it
saturates to $M^2\approx 0.0025\ \mu_B^2$ at temperatures just above $T_c$ for $x=0.28$ and 0.29 before vanishing
abruptly for $x\ge 0.30$.  The inset in Fig. 4(b) shows the P-doping dependence of the
$M^2$ above and below $T_c$ near optimal superconductivity. While superconductivity
dramatically suppresses $M^2$, it does not eliminate the ordered moment.
Figure 4(c) shows the P-doping dependence of
$\delta$ in BaFe$_2${(As$_{1-x}$P$_x$)}$_2$
below and above $T_c$.  Consistent with the P-doping dependence of $M^2$ [Fig. 4(b)] and $T_N$ [Fig. 1(c)], we find
that $\delta$ above $T_c$ approaches to $\sim 3\times 10^{-4}$ near optimal superconductivity before vanishing
at $x\ge 0.3$.

Summarizing the results in Figs. 2-4, we present the refined phase diagram of BaFe$_2${(As$_{1-x}$P$_x$)}$_2$ in Fig. 1(c).
While the present phase diagram is mostly consistent with the earlier transport and
neutron scattering work on the system
at low P-doping levels \cite{Shibauchi14,allred},
we have discovered that the magnetic and structural transitions still occur
simultaneously above $T_c$ for $x$ approaching optimal superconductivity, and
 both order parameters vanish at optimal superconductivity with $x=0.3$.  Since our NMR and TRISP measurements for
samples near optimal superconductivity suggests spin-glass-like behavior,
we conclude that the static AF order in BaFe$_2${(As$_{1-x}$P$_x$)}$_2$ disappears
in the weakly
first order fashion near optimal superconductivity.  Therefore, AF order
in phosphorus-doped iron pnictides coexists and competes
superconductivity near optimal superconductivity, much like the electron-doped iron pnictides with an avoided QCP.
From the phase diagrams of hole-doped Ba$_{1-x}A_x$Fe$_2$As$_2$ \cite{Avci2012,Avci2014,Waber,bohmer}, it appears that
a QCP may be avoided there as well.

We thank Q. Si for helpful discussions.
The work at IOP,  CAS, is supported by MOST (973 project: 2012CB821400, 2011CBA00110, and 2015CB921302) , NSFC (11374011 and 91221303) and CAS (SPRP-B: XDB07020300).  The work at Rice is supported by U.S. NSF, DMR-1362219 and by the Robert A. Welch Foundation Grants
No. C-1839. This research used resources of the APS, a User Facility operated for the DOE Office of Science
by ANL under Contract No. DE-AC02-06CH11357.  Ames Laboratory is operated for the U.S.
DOE by Iowa State University through Contract No. DE-AC02-07CH11358.

\clearpage
\appendix

\section{Supplementary information}

\textbf{Section A: Details of the neutron and X-ray scattering, and NMR experiments}

\emph{Neutron scattering experiments:}

We have aligned the $x=$ 0.19 samples in the [$H, K, 3H$] scattering plane and the $x=$ 0.25, 0.28, 0.29, 0.30, 0.31 samples in the [$H, 0, L$] zone.For neutron scattering measurements of the $x=$0.19 compound at C5 spectrometer, we used a vertically focused PG(002) monochrometor and a flat PG(002) analyzer with a fixed final energy $E_f$=14.56 meV.  We useda PG filter after the sample to eliminate the higher order neutrons.At RITA-II, we use a PG filter before the sample and a cold Be-filter after the sample with the final neutron energy fixed at 4.6 meV. For MIRA measurements, the final energy was set to $E_f=$ 4.06 meV and a Be-filter was additionally used as a filter.

In addition to usual neutron diffraction experiments, we have also carried out measurements on TRISP at MLZ, Germany.  The experimental set these measurements are described in detail in Ref. [17] of the main text.

\emph{X-ray scattering experiments:}

The high resolution X-ray diffraction of the $x=$0.28 sample was performed using a four circle diffractometer and Cu $K_{\alpha1}$ X-ray radiation from a rotating anode X-ray source at Ames Lab.  We have used beamline 6-ID-D at the Advanced Photon Source at Argonne National Laboratory with 100.2 keV incident photon beam for measurements of the $x=$ 0.19, 0.25, 0.29, 0.30, 0.31 compounds.The NMR measurements were performed by the Spin-echo technique, and the paramagnetic spectral weights were obtained by integrating the spectral intensity at the resonance frequency of the paramagnetic phase.

\textbf{Section B:additional transport, X-ray and neutron scattering data:}

We have carried careful transport measurement using 4 probe method in a physical property measurement system.  Our systematic measurements of the resistivity for different doping concentrations are shown in Fig. S1.  Typical raw data for X-ray scattering experiments is shown in Fig. S2 for the $x=$0.19 and 0.28.  The presence of two peaks along the [$H,0,0$] direction is a direct indication of lattice orthorhombicity.  Figure S3 shows the temperature dependence of the magnetic order parameter for $x=$ 0.19, 0.25, 0.28, 0.29, 0.30, 0.31 samples.  Figure S4 shows the raw $^{31}$P NMR spectra for the $x=$ 0.25 and 0.29 samples at different temperatures.

\begin{figure}[t]
\renewcommand\thefigure{S1}
\includegraphics[scale=.75]{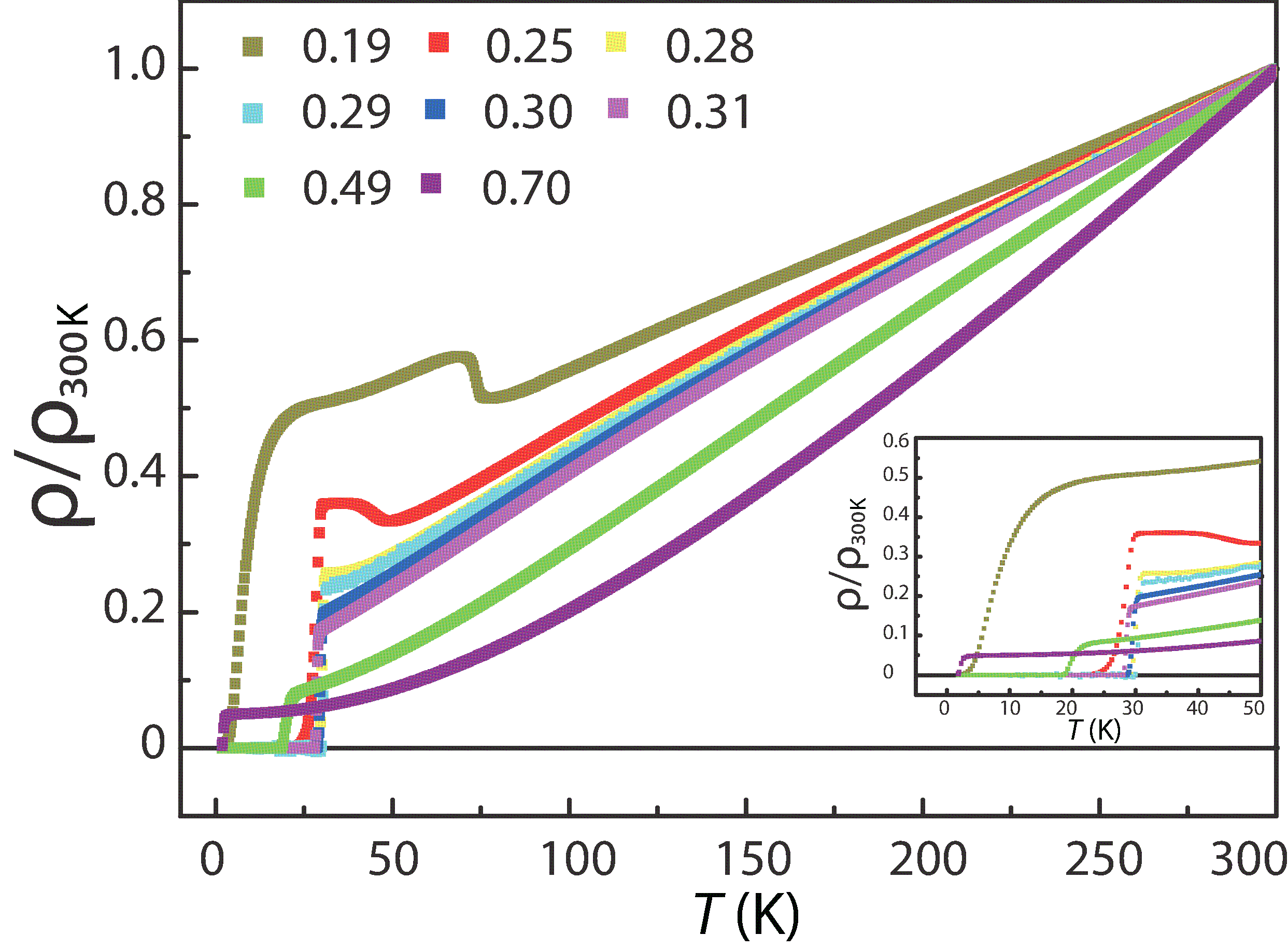}
\caption{Temperature dependence of the normalized resistivity for BaFe$_2$(As$_{1-x}$P$_x$)$_2$. The measurements were conducted by a standard four-terminal method using a Quantum Design Physical Property Measurement System. The inset shows the expanded data below 50K.
 }
\end{figure}

\begin{figure}[t]
\renewcommand\thefigure{S2}
\includegraphics[scale=.75]{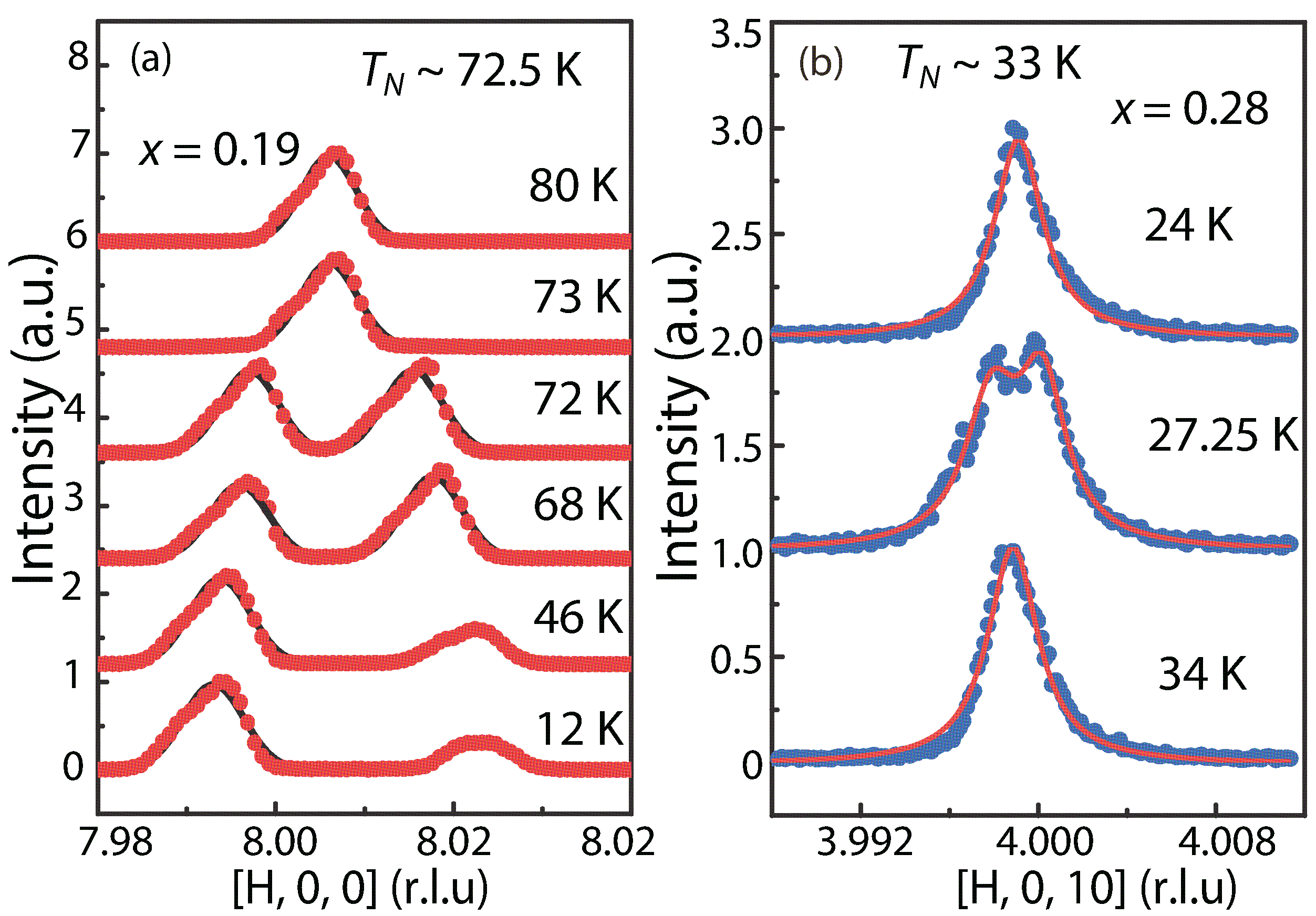}
\caption{Temperature evolution of the orthorhombic Bragg peaks for BaFe$_2$(As$_{1-x}$P$_x$)$_2$ with $x=$0.19 and 0.28. The data were collected while warming the system from the base temperature.
 }
\end{figure}

\begin{figure}[t]
\renewcommand\thefigure{S3}
\includegraphics[scale=.75]{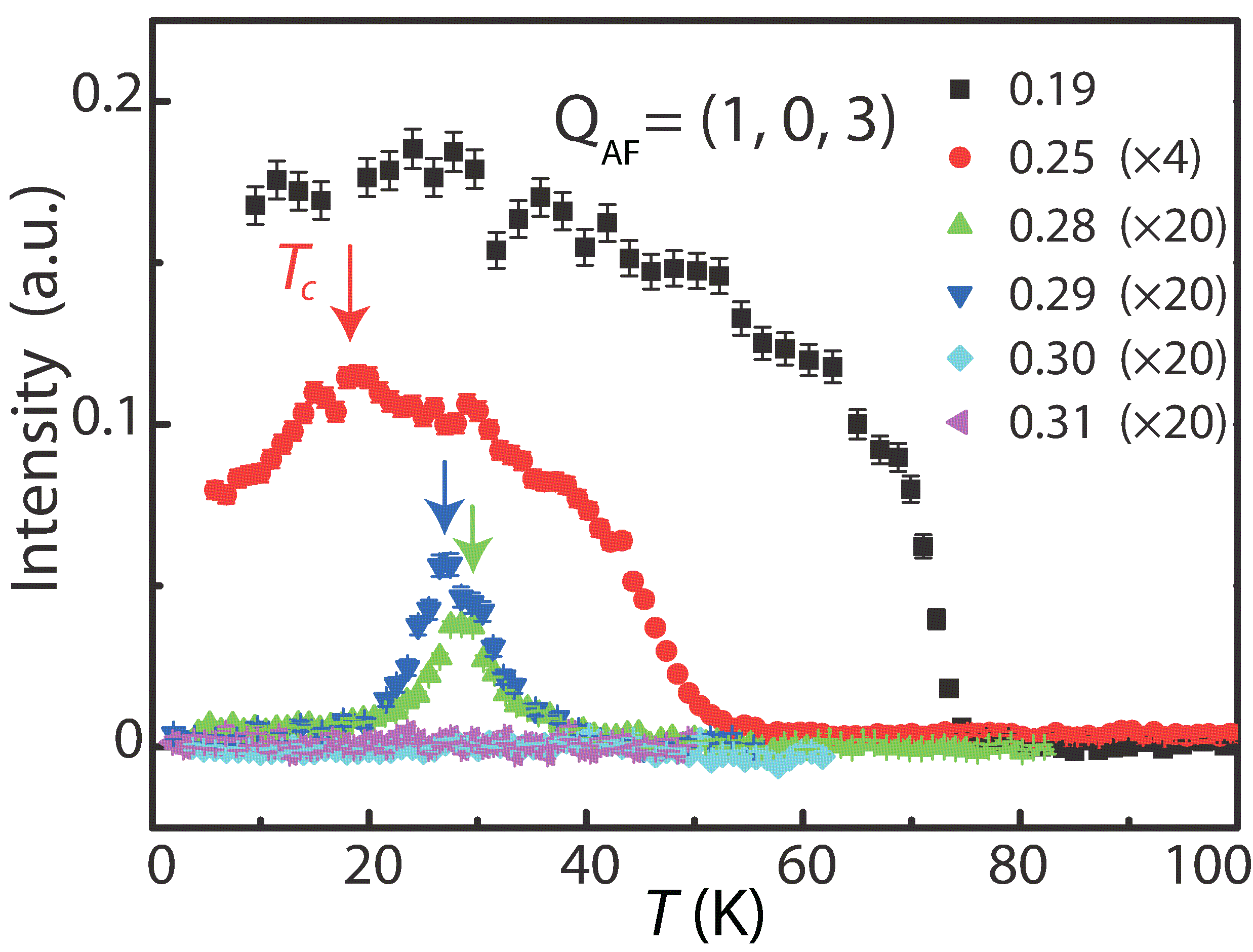}
\caption{Temperature dependence of the magnetic order parameter, where the intensity of the magnetic scattering is obtained by subtracting the data well above TN as background and normalized to weak nuclear Bragg peaks.The intensity of the $x=$ 0.25 compound was expanded by 4 and $x=$ 0.28, 0.29, 0.30, 0.31 compounds was expanded by 20. Arrows indicate the $T_c$ of different samples.The slight intensity differences for the $x=$ 0.29 and 0.30 samples are within the error of our measurements.  The observed magnetic peaks are resolution limited, giving an estimated spin-spin correlation length of ~300 \AA.
 }
\end{figure}

\begin{figure}[t]
\renewcommand\thefigure{S4}
\includegraphics[scale=.75]{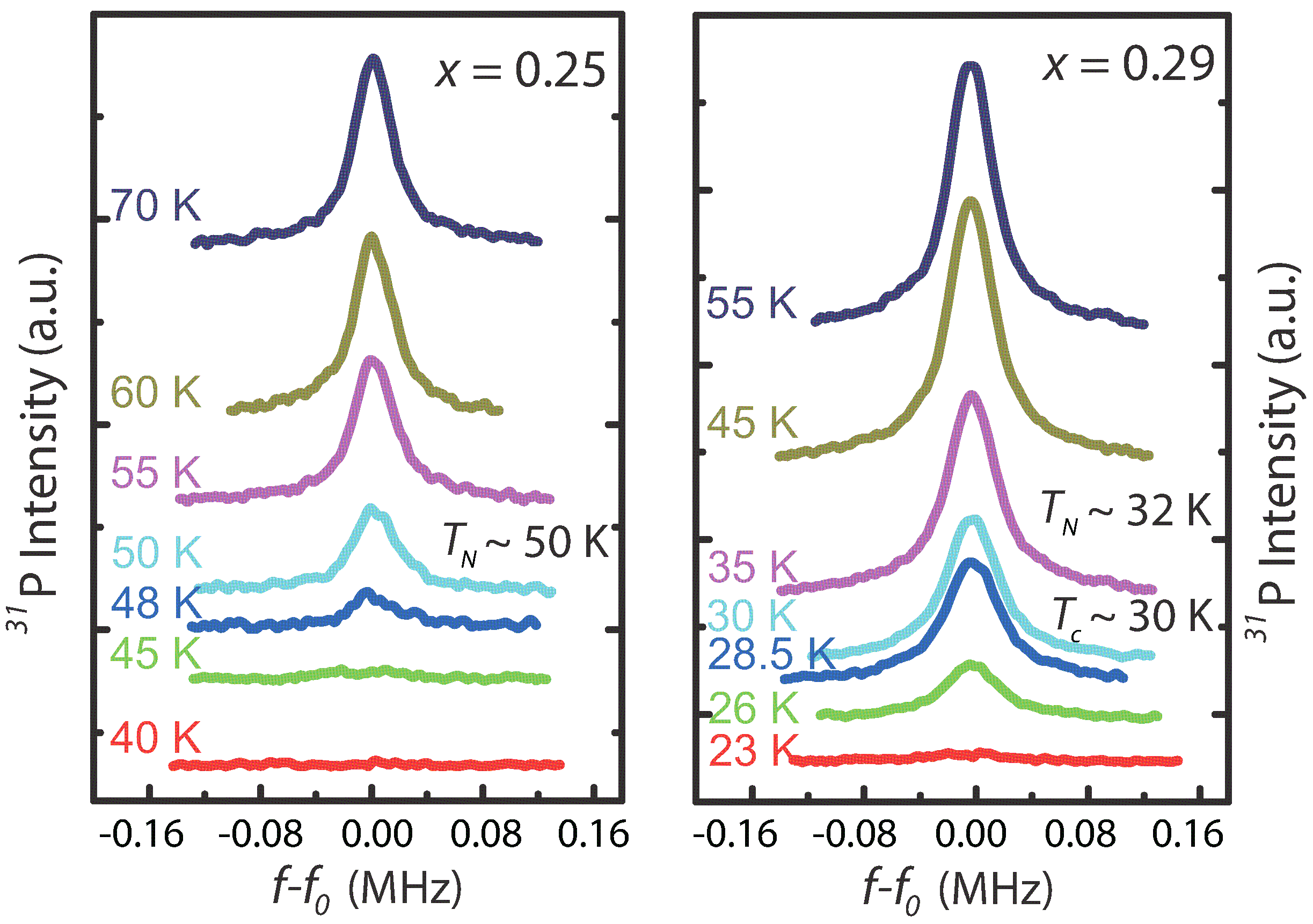}
\caption{ The $^{31}$P spectra at different temperatures for $x=$ 0.25 and $x=$ 0.29 samples. $T_N$ and $T_c$ marks the N$\acute{e}$el temperature and the superconducting transition temperature, respectively. The horizontal axes show the relative frequency from the fixed frequency $f_0$.
 }
\end{figure}


\begin{thebibliography}{}

\bibitem{kamihara} Y. Kamihara, T. Watanabe, M. Hirano, and H. Hosono,
  J. Am. Chem. Soc. \textbf{130}, 3296-3297 (2008).

\bibitem{cruz} C. de la Cruz \textit{et al.}, Nature (London) \textbf{453},899 (2008).

\bibitem{qhuang} Q. Huang, Y. Qiu, Wei Bao,
M. A. Green, J. W. Lynn, Y. C. Gasparovic, T. Wu, G. Wu, and X. H. Chen, Phys. Rev. Lett. {\bf 101}, 257003 (2008).

\bibitem{mgkim11}
M. G. Kim, R. M. Fernandes, A. Kreyssig, J. W. Kim, A. Thaler, S. L. Bud'ko,
P. C. Canfield, R. J. McQueeney, J. Schmalian, and A. I. Goldman, Phys. Rev. B {\bf 83}, 134522 (2011).

\bibitem{dai} P. Dai, J. Hu, and E. Dagotto, Nature Phys. {\bf 8}, 709 (2012).

\bibitem{Rotter} M. Rotter, M. Tegel, and D. Johrendt, Phys. Rev. Lett. {\bf 101}, 107006 (2008).

\bibitem{Gil} R. Cortes-Gil, D. R. Parker, M. J. Pitcher, J. Hadermann, and S. J.
Clarke, Chem. Mater. {\bf 22}, 4304 (2010).

\bibitem{Avci2012} S. Avci {\it et al.}, Phys. Rev. B {\bf 85}, 184507 (2012).

\bibitem{Avci2014} S. Avci {\it et al.}, Nat. Commun. {\bf 5}, 3845 (2014).

\bibitem{Waber} F. Wa$\rm \beta$er {\it et al.}, Phys. Rev. B {\bf 91}, 060505(R) (2015).

\bibitem{bohmer} A. E. B$\rm \ddot{o}$hmer, F. Hardy, L. Wang, T. Wolf, P. Schweiss, and
C. Meingast, arXiv: 1412.7038v2.

\bibitem{canfield} P. C. Canfield and S. L. Bud'ko, Annu. Rev. Condens. Matter Phys. {\bf 1}, 27 (2010).

\bibitem{fisher} I. R. Fisher, L. Degiorgi, and Z.-X. Shen, Rep. Prog. Phys. {\bf 74}, 124506 (2011).

\bibitem{bernhard} C. Bernhard, C. N. Wang, L. Nuccio, L. Schulz, O. Zaharko,
J. Larsen, C. Aristizabal, M. Willis, A. J. Drew, G. D.
Varma, T. Wolf, and C. Niedermayer, Phys. Rev. B {\bf 86}, 184509 (2012).

\bibitem{ning} F. L. Ning, K. Ahilan, T. Imai, A. S. Sefat, M. A. McGuire,
B. C. Sales, D. Mandrus, P. Cheng, B. Shen, and H.-H.
Wen, Phys. Rev. Lett. {\bf 104}, 037001 (2010).

\bibitem{zhou2013} R. Zhou, Z. Li, J. Yang, D. L. Sun, C. T. Lin, and G.-Q.
Zheng, Nat. Commun. {\bf 4}, 2265 (2013).

\bibitem{dioguardi} A. P. Dioguardi {\it et al.},
Phys. Rev. Lett. {\bf 111}, 207201 (2013).

\bibitem{clester} C. Lester, J.-H. Chu, J. G. Analytis, S. C. Capelli, A. S. Erickson, C. L. Condron, M. F. Toney, I. R. Fisher, and S. M. Hayden,
Phys. Rev. B {\bf 79}, 144523 (2009).

\bibitem{nandi} S. Nandi {\it et al.}, Phys. Rev. Lett. {\bf 104}, 057006 (2010).

\bibitem{dkpratt11} D. K. Pratt {\it et al.}, Phys. Rev. Lett. {\bf 106}, 257001 (2011).

\bibitem{hqluo} H. Luo {\it et al.},
 Phys. Rev. Lett. {\bf 108}, 247002 (2012).

\bibitem{xylu} X. Y. Lu {\it et al.}, Phys. Rev. Lett. {\bf 110}, 257001 (2013).

\bibitem{xylu14} X. Y. Lu {\it et al.}, Phys. Rev. B {\bf 90}, 024509 (2014).

\bibitem{eabrahams11} E. Abrahams and Q. Si, J. Phys. Condens. Matter {\bf 23}, 223201 (2011).

\bibitem{SJiang} S. Jiang, C. Wang, Z. Ren, Y. Luo, G. Cao, and Z.-A. Xu,
J. Phys. Condens. Matter {\bf 21}, 382203 (2009).


\bibitem{Shishido10} H. Shishido {\it et al.}, Phys. Rev. Lett. {\bf 104}, 057008 (2010).

\bibitem{Beek10} C. J. van der Beek, M. Konczykowski, S. Kasahara,
T. Terashima, R. Okazaki, T. Shibauchi, and Y. Matsuda,
Phys. Rev. Lett. {\bf 105}, 267002 (2010).

\bibitem{Kasahara10} S. Kasahara, T. Shibauchi, K. Hashimoto, K. Ikada,
S. Tonegawa, R. Okazaki, H. Shishido, H. Ikeda, H.
Takeya, K. Hirata, T. Terashima, and Y. Matsuda, Phys. Rev. B {\bf 81}, 184519 (2010).

\bibitem{Hashimoto12} K. Hashimoto {\it et al.}, Science {\bf 336}, 1554 (2012).

\bibitem{Shibauchi14} T. Shibauchi, A. Carrington, and Y. Matsuda, Annu.
Rev. Condens. Matter Phys. {\bf 5}, 113 (2014).

\bibitem{Walmsley} P. Walmsley {\it et al.}, Phys. Rev. Lett. {\bf 110}, 257002 (2013).

\bibitem{Analytis14} J. G. Analytis, H.-H. Kuo, R. D. McDonald, M. Wartenbe, P. M. C. Rourke,
N. E. Hussey, and I. R. Fisher, Nat. Phys. {\bf 10}, 194 (2014).

\bibitem{Nakai10} Y. Nakai, T. Iye, S. Kitagawa, K. Ishida, H. Ikeda, S.
Kasahara, H. Shishido, T. Shibauchi, Y. Matsuda, and T.
Terashima, Phys. Rev. Lett. {\bf 105}, 107003 (2010).


\bibitem{allred} J. M. Allred {\it et al.}, Phys. Rev. B {\bf 90}, 104513 (2014).


\bibitem{keller} T. Keller, K. Habicht, H. Klann, M. Ohl, H. Schneier, and
B. Keimer, Appl. Phys. A {\bf 74}, s332 (2002).

\bibitem{supplementary} See supplemental Material for a
detailed discussion on the experimental setup and additional data.

\bibitem{sample} M. Nakajima, S. Uchida, K. Kihou, C. H. Lee, A. Iyo, and H. Eisaki,
J. Phys. Soc. Jpn. {\bf 81}, 104710 (2012).

\end{thebibliography}
\end{document}